\documentclass[pra,twocolumn,superscriptaddress]{revtex4-1}
\usepackage{amsmath}
\usepackage{graphicx}
\usepackage{theorem}

\theorembodyfont{\normalfont}
\newtheorem{theorem}{Theorem}

\newtheorem{conjecture}{Conjecture}

\begin{document}
\begin{flushright}
  YITP-19-11
\end{flushright}   

\title{Fine-grained quantum supremacy based on Orthogonal Vectors, 3-SUM and All-Pairs Shortest Paths}
\author{Ryu Hayakawa}
\email{ryu.hayakawa@yukawa.kyoto-u.ac.jp}
\affiliation{Yukawa Institute for Theoretical Physics,
Kyoto University, Kitashirakawa Oiwakecho, Sakyoku, Kyoto
606-8502, Japan}
\author{Tomoyuki Morimae}
\email{tomoyuki.morimae@yukawa.kyoto-u.ac.jp}
\affiliation{Yukawa Institute for Theoretical Physics,
Kyoto University, Kitashirakawa Oiwakecho, Sakyoku, Kyoto
606-8502, Japan}
\affiliation{JST, PRESTO, 4-1-8 Honcho, Kawaguchi, Saitama,
332-0012, Japan}
\author{Suguru Tamaki}
\email{tamak@sis.u-hyogo.ac.jp}
\affiliation{School of Social Information Science, University of Hyogo, Japan}
\date{\today}

\begin{abstract}
Fine-grained quantum supremacy is a study of proving (nearly) tight
time lower bounds for
classical simulations of quantum computing under ``fine-grained
complexity" assumptions.
We show that under conjectures on
Orthogonal Vectors (OV), 3-SUM, 
All-Pairs Shortest Paths (APSP) and their variants,
strong and weak classical simulations of
quantum computing are impossible in certain exponential 
time 
with respect to 
the number of qubits.
Those conjectures are 
widely used 
in classical fine-grained complexity theory 
in which polynomial time hardness is conjectured.
All previous results
of fine-grained quantum supremacy
are based on ETH, SETH, or their
variants
that are conjectures for SAT 
in which exponential time hardness is conjectured.
We show that there exist quantum circuits 
which cannot be classically simulated 
in certain exponential time with respect to the number of qubits 
first by considering a Quantum Random Access Memory (QRAM) based 
quantum computing model
and next by considering a non-QRAM model quantum computation. 
In the case of the QRAM model, 
the size of quantum circuits 
is linear with respect to the number of qubits 
and in the case of the non-QRAM model, 
the size of the quantum circuits 
is exponential with respect to the number of qubits but the results 
are still non-trivial.
\end{abstract}
\maketitle

\section{Introduction}
Quantum computing is believed to have advantages in its computing time
over classical computing and
there are several approaches to show these advantages.
One way is to show that a quantum algorithm can
solve a problem faster than the best known classical algorithm,
such as Shor's factoring algorithm~\cite{Shor}.
However,
the best classical algorithm could be updated~\cite{Tang}.
Another approach is based on query complexity,
which means to evaluate the number of times to
call a certain subroutine.
Grover's search algorithm~\cite{Grover} is a representative
of this kind of approach.
In query complexity,
the advantage can be unconditionally proven
but
we do not know about the real time of computation.

The third approach, which has been actively studied recently, is to consider sampling problems.
It is known that output probability distributions of several sub-universal quantum computing models
cannot be classically sampled in polynomial time
within a multiplicative error
$\epsilon<1$ unless the polynomial-time
hierarchy collapses to the second level.
Here, we say that a probability distribution 
$\{p_z\}_z$ is classically sampled 
in time $T$ within a multiplicative
error $\epsilon$ if there exists a $T$-time 
classical probabilistic algorithm 
that outputs $z$ with probability $q_z$ such that
$|p_z-q_z| \le \epsilon p_z$
for all $z$.
Classically sampling output probability distributions of quantum computing
is also called a weak simulation.
In contrast,
calculating output probability distributions of quantum computing
is called a strong simulation.

Several sub-universal models that exhibit such ``quantum supremacy"
have been found such as
the depth-four model~\cite{TD},
the Boson Sampling model~\cite{BS},
the IQP model~\cite{IQP,IQP2},
the one-clean-qubit model~\cite{KL,MFF,FKMNTT,FKMNTT2},
the random circuit model~\cite{randomcircuit,Movassagh1,Movassagh2}, 
and the HC1Q model~\cite{HC1Q}.

All these quantum supremacy results, however, prohibit
only polynomial-time classical simulations: these models could
be classically simulated in exponential time.
To show (nearly) tight time lower bounds for classical simulations of quantum computing, 
the study of more ``fine-grained" quantum supremacy has been started. 
In Ref.~\cite{Huang,Huang2},
impossibilities of some exponential-time
strong simulations were shown based on
the exponential-time hypothesis (ETH)
and the strong exponential-time hypothesis
(SETH)~\cite{SETH1,SETH2,SETH3}.
Ref.~\cite{Dalzell,Dalzell2} showed that output 
probabilities of
the IQP model, the QAOA model~\cite{QAOA}, and
the Boson Sampling model cannot be classically sampled
in some exponential time within a multiplicative error $\epsilon<1$
under some SETH-like conjectures.
Ref.~\cite{MT} showed similar results for
the one-clean-qubit model and the HC1Q model.
Refs.~\cite{MT,Huang2} also studied fine-grained quantum supremacy
of Clifford-$T$
quantum computing, and Ref.~\cite{MT} studied
Hadamard-classical quantum computing.

All previous results~\cite{Huang,Huang2,Dalzell,Dalzell2,MT}
on fine-grained quantum supremacy
are based on ETH, SETH, or their
variants
in which exponential time hardness for SAT problems is conjectured.  
In this paper, we show
fine-grained quantum supremacy results (in terms of
the qubit-scaling) based on
Orthogonal Vectors (OV)~\cite{WilliamsSETHOV}, 
3-SUM~\cite{3-SUM}, 
All-Pairs Shortest Paths (APSP)~\cite{WilliamsAPSPNWT} 
and their variants. 
Those are  
widely used conjectures 
in fine-grained complexity
and many reductions from those conjectures 
to other conjectures are known~\cite{WilliamsFGcomplexity}.
(There is no known reduction among those three conjectures.)
APSP is known to be equivalent to Negative Weight Triangle (NWT)~\cite{WilliamsAPSPNWT},
and therefore we use the conjecture of NWT to show 
fine-grained quantum supremacy instead of that of APSP.
Of those three conjectures,
only OV is known to be reduced from SETH~\cite{WilliamsSETHOV}.

For each conjecture, 
we first show fine-grained quantum supremacy results 
in the case when
the Quantum Random Access Memory (QRAM)~\cite{QRAM} 
is available. 
The QRAM is the quantum version of the Random Access Memory (RAM) and 
it can return a superposition of data in a single step as
\begin{eqnarray*}
\sum_i a_i|i\rangle \otimes |0^d\rangle 
\xrightarrow{QRAM}
\sum_i a_i|i\rangle \otimes |D[i]\rangle,
\end{eqnarray*}
where $D[i]$ is the $d$-bit data stored in the memory of index $i$. 
Next,
we show fine-grained quantum supremacy results 
of quantum circuits without the QRAM 
by constructing specific unitary operations
which correspond to the QRAM operations. 

The reason why we consider the QRAM model is that fine-grained complexity conjectures are usually defined with the word RAM model, and its natural correspondence seems to be the QRAM model. 
We, however, also consider the non-QRAM model as well, because the QRAM model cannot be directly realized in real experiments.

In both cases,
we show that there exist quantum circuits 
whose output probability distributions 
cannot be classically sampled 
in certain exponential time 
in terms of the number of qubits. 
In the case of the QRAM based quantum computing,  
the size of the quantum circuits is linear with respect to the number of qubits
and 
in the case of the non-QRAM model,  
the size of the quantum circuits is exponential 
with respect to the number of qubits 
but the results are still non-trivial. 

Note that when we consider ETH or SETH like conjectures, we can construct efficient quantum circuits without the QRAM, because there are no data to be stored in QRAM.

Throughout this paper, we use the
following notations.
When a non-negative integer $a$ can be written as 
\begin{eqnarray}
\label{eq:binaryrep}
a=\sum_{j=0}^{r-1}2^j a_j,
\end{eqnarray}
where $a_j \in \{0,1\}$ for $j=0,1,...,r-1$.
We define its $r$-bit binary representation as
\begin{eqnarray}
B[a] \equiv (a_0,a_1,...,a_{r-1}) \in \{0,1\}^r.
\end{eqnarray}
Also, 
when we have an $r$-bit string $x=(x_0,x_1,...,x_{r-1})$, 
we define its integer representation as 
\begin{eqnarray}
\label{eq:integerrep}
I[x]\equiv \sum_{j=0}^{r-1}2^j x_j.
\end{eqnarray}
Let $a=(a_0,a_1,...,a_{r-1})$ be an $r$-bit string.
We define
\begin{eqnarray}
\label{eq:Xbinary}
\begin{aligned}
X^{a}&\equiv \bigotimes_{j=0}^{r-1} X^{a_j},\\
X^{a\oplus 1}&\equiv \bigotimes_{j=0}^{r-1} X^{a_j\oplus 1},
\end{aligned}
\end{eqnarray}
where $X$ is the Pauli-$X$ operator.
Let us denote the $d$-qubit-controlled $X^a$ gate as
$\Lambda_{d}(X^{a})$, 
which acts as
\begin{eqnarray}
\label{eq:controlledgate}
&&\Lambda_{d}(X^{a})|x_0,x_1,...,x_{d-1}\rangle\otimes
|y_0,y_1,...,y_{r-1}\rangle\\ 
&&=\left\{
    \begin{array}{l}
    |x_0,x_1,...,x_{d-1}\rangle\otimes
    |y_0\oplus a_0,y_1\oplus a_1,...,y_{r-1}\oplus a_{r-1}\rangle\\
    \hspace{3.2cm} ({\rm if}\ x_0=x_1=\cdots=x_{d-1}=1),\\
    |x_0,x_1,...,x_{d-1}\rangle\otimes
    |y_0,y_1,...,y_{r-1}\rangle
    \hspace{0.8cm} ({\rm otherwise}),
    \end{array}
\right. \nonumber
\end{eqnarray}
for all $(x_0,x_1,...,x_{d-1})\in \{0,1\}^d$ 
and $(y_0,y_1,...,y_{r-1})\in\{0,1\}^r$. 
$\Lambda_{d}(X^{a})$ can be composed of 
$r$-number of $d$-controlled TOFFOLI gates (generalized TOFFOLI gates). 
A $d$-controlled TOFFOLI gate
can be decomposed into $8(d-3)$-number of TOFFOLI gates with a single ancilla qubit that can be reused without
any initialization 
as it is shown in the Corollary 7.4 of Ref.~\cite{Barenco}.

There are quantum circuits that can compare two binary integers. 
In Appendix~\ref{app:judge}, 
we construct a quantum circuit $C$ such that 
\begin{eqnarray}
&&C(|0\rangle \otimes |a_0,a_1,...,a_{r-1}\rangle \otimes |b_0,b_1,...,b_{r-1}
\rangle \otimes |0\rangle )\nonumber\\
&&
\begin{aligned}
=|0\rangle \otimes |a_0,a_1,...,a_{r-1}\rangle \otimes |b_0,&b_1,...,b_{r-1}
\rangle\nonumber& \\
&\otimes |\chi(I(a)-I(b))\rangle,&
\end{aligned}
\label{eq:C}
\end{eqnarray}
where
\begin{eqnarray}
\label{eq:chi}
\chi(x)=
  \left\{
    \begin{array}{l}
      0 \ (x\leq 0), \\
      1 \ (x> 0).
    \end{array}
  \right.
\end{eqnarray}
We also construct a quantum circuit $C'$ such that
\begin{eqnarray}
\label{eq:Cprime}
&&C'(|0\rangle \otimes |a_0,a_1,...,a_{r-1}\rangle \otimes |b_0,b_1,...,b_{r-1}
\rangle \otimes |0\rangle )\\
&&\ 
\begin{aligned}
=|0\rangle \otimes |a_0,a_1,...,a_{r-1}\rangle \otimes |b_0&,b_1,...,b_{r-1}
\rangle& \nonumber \\
&\otimes |\chi(I(b)-I(a))\oplus 1\rangle.&
\end{aligned}
\end{eqnarray}
Note that the quantum circuit $C$ decides whether $I[a]\leq I[b]$ or not 
while $C'$ does whether $I[a]<I[b]$ or not. 
(For details, see Appendix~\ref{app:judge}.) 

There are quantum circuits that can do the addition.
For example, in Ref.~\cite{addition}, the circuit $A$ 
was introduced such that
\begin{eqnarray}
\label{eq:addition}
\begin{aligned}
A(|0\rangle\otimes|a_0,...,a_{r-1}
\rangle\otimes|b_0,...,b_{r-1}\rangle\otimes|0\rangle)\\
=|0\rangle\otimes|a_0,...,a_{r-1}\rangle
\otimes|s_0,...,s_{r-1}\rangle\otimes|s_r\rangle
\end{aligned}
\end{eqnarray}
for any non-negative $r$-bit strings $a$
and $b$, 
and
$a+b=\sum_{j=0}^r2^js_j$
with $(s_0,...,s_r)\in\{0,1\}^{r+1}$. 
(For details, see Appendix~\ref{app:addition}.)

\section{Orthogonal Vectors}
\label{sec:ov}
In this section, 
we show fine-grained quantum supremacy 
in terms of the qubit scaling 
based on Orthogonal Vectors and its variant. 
Let us introduce the following two conjectures:

\begin{conjecture}[Orthogonal Vectors]
\label{conjecture:OV}
For any $\delta>0$, there is a $c$ such that
deciding whether $s>0$ or $s=0$
for given vectors, 
$u_1,...,u_n,v_1,...,v_n\in\{0,1\}^d$,
with $d=c\log n$
cannot be done in time $n^{2-\delta}$. 
Here 
\begin{eqnarray*}
s\equiv|\{(i,j)~|~u_i\cdot v_j=0\}|.
\end{eqnarray*}
\end{conjecture}

\begin{conjecture}
\label{conjecture:OVgapNTIME}
For any $\delta>0$, there is a $c$ such that
deciding whether $gap\neq0$ or $gap=0$
for given vectors, 
$u_1,...,u_n,v_1,...,v_n\in\{0,1\}^d$,
with $d=c\log n$
cannot be done in non-deterministic time $n^{2-\delta}$. 
Here, 
\begin{eqnarray*}
gap\equiv|\{(i,j)~|~u_i\cdot v_j=0\}|
-|\{(i,j)~|~u_i\cdot v_j\neq0\}|.
\end{eqnarray*}
\end{conjecture}

We use two different acceptance criteria, 
one is on \#P functions, 
which is usually considered in fine-grained complexity theory, 
and the other is on gap functions. 
The conjecture on gap functions is also justified 
because the only known way to decide whether 
$gap\neq 0$ or $gap=0$ is to solve \#P problems. 
The same can be said to the conjectures in the later sections. 

Thinking of the QRAM model quantum computing, 
we can show the following two results 
based on the above two conjectures:

\begin{theorem}[Strong simulation with QRAM]
\label{theorem:ssOVqram}
Assume that Conjecture~\ref{conjecture:OV} is true. Then,
for any $\delta>0$, there is a $c$ such that
there exists an $N$-qubit and $\mathcal{O}(N)$-size  
quantum circuit with access to the QRAM 
whose acceptance probability 
cannot be classically exactly calculated
in time $T\equiv 2^{\frac{(2-\delta)(N-7)}{3(c+1)}}$.
\end{theorem}

\begin{theorem}[Weak simulation with QRAM]
\label{theorem:wsOVqram}
Assume that Conjecture~\ref{conjecture:OVgapNTIME}
is true. Then,
for any $\delta>0$, there is a $c$ such that
there exists an $N$-qubit and $\mathcal{O}(N)$-size 
quantum circuit with access to the QRAM whose 
acceptance probability 
cannot be classically sampled within a multiplicative error $\epsilon<1$
in time $T\equiv 2^{\frac{(2-\delta)(N-7)}{3(c+1)}}$.
\end{theorem}
    
By constructing a unitary operation 
corresponding to the QRAM process, 
we can show the following two results 
based on the above two conjectures:

\begin{theorem}[Strong simulation]
\label{theorem:ssOV}
Assume that Conjecture~\ref{conjecture:OV} is true. Then,
for any $\delta>0$, there is a $c$ such that
there exists an $N$-qubit 
and $\mathcal{O}(N^22^\frac{N}{3(c+1)})$-size 
quantum circuit whose 
acceptance probability 
cannot be classically exactly calculated
in time $T\equiv 2^{\frac{(2-\delta)(N-7)}{3(c+1)}}$.
\end{theorem}
    
\begin{theorem}[Weak simulation]
\label{theorem:wsOV}
Assume that Conjecture~\ref{conjecture:OVgapNTIME}
is true. Then,
for any $\delta>0$, there is a $c$ such that
there exists an $N$-qubit 
and $\mathcal{O}(N^22^{\frac{N}{3(c+1)}})$-size 
quantum circuit whose 
acceptance probability 
cannot be classically sampled within a multiplicative error $\epsilon<1$
in time $T\equiv 2^{\frac{(2-\delta)(N-7)}{3(c+1)}}$.
\end{theorem}

{\it Proof of Theorem \ref{theorem:ssOVqram} and \ref{theorem:wsOVqram}}. 
For given $n$, 
let $r$ be the smallest integer such that $n\leq 2^r$, i.e.,
\begin{eqnarray}
  \label{eq:rofOV}
  \begin{aligned}
  2^{r-1}< &n \leq 2^r \\
  \Leftrightarrow \log_2{n}\leq &r < \log_2{n}+1.
  \end{aligned}
\end{eqnarray}
For given vectors $u_1,...,u_n,v_1,...,v_n\in\{0,1\}^d$, 
we can think of the QRAM which stores the data of those vectors as 
\begin{eqnarray}
  \label{eq:ovdata}
  \begin{aligned}
D[i]=u_{I[i]+1}\in\{0,1\}^d,\\
D'[j]=v_{I[j]+1}\in\{0,1\}^d,
  \end{aligned}
\end{eqnarray}
for $i,j\in\{B[0],B[1],...,B[n-1]\}$.

Let us consider the following 
quantum computing:

\begin{itemize}
\item[1.]
Generate
\begin{eqnarray*}
\frac{1}{2^r}
\sum_{i,j\in\{0,1\}^r}|i\rangle_1\otimes |j\rangle_2 \otimes |B[n-1]\rangle_3
\otimes |00\rangle_4 \\
\otimes |0^d\rangle_5 \otimes |0^d\rangle_6 
\otimes |0^d\rangle_7 
\otimes |0\rangle_8.
\end{eqnarray*}
We have introduced subscript numbers which represent the indices of registers.

\item[2.]
Apply the quantum circuit $C$ of Eq.~(\ref{eq:C}) 
between the 1st-3rd registers and between the 2nd-3rd registers, 
and flip the first and second qubits of the 4th register 
according to their results, 
respectively. 
Then we get 
\begin{eqnarray*}
&&\frac{1}{2^r}
\sum_{i,j\in\{0,1\}^r}|i\rangle_1\otimes |j\rangle_2
\otimes |B[n-1]\rangle_3 \\
&&\ \otimes 
|\chi (I[i]-n+1),\chi (I[j]-n+1)\rangle_4 \\
&&\ \otimes |0^d\rangle_5 \otimes |0^d\rangle_6 \otimes |0^d\rangle_7 \otimes |0\rangle_8.
\end{eqnarray*}
Note that $|\chi (I[i]+1-n),\chi (I[j]+1-n)\rangle$ is $|00\rangle$ 
if $I[i]+1\in\{1,2,...,n\}$ and $I[j]+1\in\{1,2,...,n\}$. 

\item[3.]
Access to the QRAM using
the first register as the address of $D$ 
and the second register as the address of $D'$ 
and map the results to the 5th register and the 6th register, 
respectively.
For $i$ and $j$ which are larger than $n-1$ ($n-1 < I[i], I[j] \leq 2^r-1$), 
there are no data of $D[i]$ and $D'[j]$, 
then we assume the registers of $|D[i]\rangle$ and $|D'[j]\rangle$ are $|0^d\rangle$ 
for such $i$ and $j$. 
Then we get 
\begin{eqnarray*}
  &&\frac{1}{2^r}
  \sum_{i,j\in\{0,1\}^r}|i\rangle_1\otimes |j\rangle_2
  \otimes |B[n-1]\rangle_3 \\
  &&\ \otimes 
  |\chi (I[i]-n+1),\chi (I[j]-n+1)\rangle_4 \\
  &&\ \otimes |D[i]\rangle_5 \otimes |D'[j]\rangle_6 \otimes |0^d\rangle_7 \otimes |0\rangle_8.
\end{eqnarray*}

\item[4.]
Apply bit-wise TOFFOLI on the 5th,
6th, and 7th registers to generate
\begin{eqnarray*}
&&\frac{1}{2^r}
\sum_{i,j\in\{0,1\}^r}|i\rangle_1\otimes |j\rangle_2
\otimes |B[n-1]\rangle_3 \\
&&\ \otimes 
|\chi (I[i]-n+1),\chi (I[j]-n+1)\rangle_4 \\
&&\ \otimes |D[i]\rangle_5 \otimes |D'[j]\rangle_6 \otimes |D[i]\cdot D'[j]\rangle_7 \otimes |0\rangle_8,
\end{eqnarray*}
where
$
D[i]\cdot D'[j] =
(D[i]_1D'[j]_1,...,D[i]_dD'[j]_d)
$
.
\item[5.]
Flip the 8th register if and only if the 7th register is
$|0^d\rangle$:
\begin{eqnarray*}
  &&\frac{1}{2^r}
  \sum_{i,j\in\{0,1\}^r}|i\rangle_1\otimes |j\rangle_2
  \otimes |B[n-1]\rangle_3 \\
  &&\ \otimes 
  |\chi (I[i]-n+1),\chi (I[j]-n+1)\rangle_4 \\
  &&\ \otimes |D[i]\rangle_5 \otimes |D'[j]\rangle_6 \otimes |D[i]\cdot D'[j]\rangle_7 \otimes |\delta_{D[i]\cdot D'[j],0^d}\rangle_8.
\end{eqnarray*}
This can be done by applying 
\begin{eqnarray*}
(X^{\otimes d}\otimes I)\cdot(\Lambda_d(X))\cdot(X^{\otimes d}\otimes I)
\end{eqnarray*}
between the 7th-8th registers, where $\Lambda_d(X)$ 
is the $d$-controlled $X$ gate defined in Eq.~(\ref{eq:controlledgate}). 

\item[6.]
Apply $Z$ gate to the last qubit 
and finally get
\begin{eqnarray*}
  &&\frac{1}{2^r}
  \sum_{i,j\in\{0,1\}^r}
  (-1)^{\delta_{D[i]\cdot D'[j],0^d}}
  |i\rangle_1\otimes |j\rangle_2
  \otimes |B[n-1]\rangle_3 \\
  &&\ \otimes 
  |\chi (I[i]-n+1),\chi (I[j]-n+1)\rangle_4 \\
  &&\ \otimes |D[i]\rangle_5 \otimes |D'[j]\rangle_6 \otimes |D[i]\cdot D'[j]\rangle_7 \otimes |\delta_{D[i]\cdot D'[j],0^d}\rangle_8\\
&&\hspace{0.5cm}\equiv |\Phi\rangle.
\end{eqnarray*}

\item[7.]
Measure qubits of the 4th register of $|\Phi\rangle$ 
in the $Z$ basis
and 
measure all the other qubits of $|\Phi\rangle$ 
in the $X$ basis. 
If all results are $0$, 
then accept. 
Then, 
the acceptance probability is 
\begin{eqnarray}
\label{eq:paccOV}
p_{acc}
\equiv
|\langle+^{3r}00+^{3d}+|\Phi\rangle|^2=
\frac{gap^2}{2^{5r+3d+1}},
\end{eqnarray}
where $|+\rangle = (|0\rangle + |1\rangle)/ \sqrt{2}$. 
\end{itemize}

This quantum computing needs $3d+3r+4$ qubits.
The reason is as follows: first,
it is clear that $3d+3r+3$ qubits are needed.
Second, each of the quantum circuit $C$ and the generalized TOFFOLI gate used in the above quantum computing 
needs a single ancilla qubit which can be reused without initialization.
Hence we only need a single ancilla qubit for these quantum circuits.
Thus in total, $3d+3r+4\equiv N$ qubits are necessary. Then the following inequality holds using Eq.~(\ref{eq:rofOV}):
\begin{eqnarray*}
N=3d+3r+4&&< 3c\log_2{n}+3(\log_2{n}+1)+4\\
&&= 3(c+1)\log_2{n}+7.
\end{eqnarray*}

We summarize the number of quantum gates used at most
in each step in table~\ref{tab:sizeOV}. 
(`At most' means that, for example,  
we need $r$ number of $X$-gates to generate 
$|B[n-1]\rangle$ from $|0^r\rangle$ in step 1 
if $B[n-1]=1^r$ and 
we need less if not.)
As it can be seen from this table, 
this quantum computing uses $\mathcal{O}(N)$ quantum gates. 

\begin{table}[thb]
  \caption{The number of quantum gates 
  used at most in each step of the quantum computation
  of OV. $CX$-gate means the controlled-Pauli $X$ gate.}
  \begin{tabular}{|c|c|r|} \hline
    step & gate &number \\ \hline \hline
    1.   & $H$-gate  & $2r$     \\ 
         & $X$-gate  & $r$     \\ \hline
    2.   & $X$-gate  & $4r+6$     \\ 
         & $CX$-gate  & $8r+2$     \\
         & TOFFOLI  & $4r$     \\ \hline
    3.   & QRAM     &  $2$   \\ \hline
    4.   & TOFFOLI  & $d$     \\ \hline
    5.   & $X$-gate & $2d$ \\
         & TOFFOLI  & $8(d-3)$     \\ \hline
    6.   & $Z$-gate      & 1 \\ \hline
   Non-QRAM 
         & $X$-gate      & $4nr$ \\ 
   unitary operation      & TOFFOLI       & $16nd(r-3)$ \\ \hline
  \end{tabular}
  \label{tab:sizeOV}
\end{table}

Let us define $T$ as 
\begin{eqnarray*}
T\equiv 2^{\frac{(2-\delta)(N-7)}{3(c+1)}}<
n^{2-\delta}.
\end{eqnarray*}
Assume that $p_{acc}$ of Eq.~(\ref{eq:paccOV}) can be classically exactly calculated in time $T$.
Then, $|\{(i,j)~|~u_i\cdot v_j=0\}|=(gap + n^2)/2>0$ or $=0$ can be decided in
time $n^{2-\delta}$, 
which 
contradicts to Conjecture~\ref{conjecture:OV}.
Hence Theorem~\ref{theorem:ssOVqram} has been shown.
Next assume that
$p_{acc}$ can be classically sampled within a multiplicative error
$\epsilon<1$ in time $T$,  
which means that there exists a classical probabilistic 
$T$-time
algorithm that accepts with probability $q_{acc}$ such that
\begin{eqnarray*}
|p_{acc}-q_{acc}|\le\epsilon p_{acc}.
\end{eqnarray*}
If $gap\neq0$, then
\begin{eqnarray*}
q_{acc}\ge(1-\epsilon)p_{acc}>0.
\end{eqnarray*}
If $gap=0$, then
\begin{eqnarray*}
q_{acc}\le(1+\epsilon)p_{acc}=0.
\end{eqnarray*}
It means that deciding $gap\neq0$ or $gap=0$
can be done in non-deterministic time
$n^{2-\delta}$, which
contradicts to 
Conjecture~\ref{conjecture:OVgapNTIME}.
Hence Theorem~\ref{theorem:wsOVqram} has been shown.
\fbox

\vspace{0.3cm}
{\it Proof of Theorem \ref{theorem:ssOV} and \ref{theorem:wsOV}}.
This can be done by just replacing the QRAM operation of the above proof 
by a specific unitary operation. 
For the data
\begin{eqnarray*}
  \begin{aligned}
D[i]=u_{I[i]+1}\in\{0,1\}^d,\\
D'[j]=v_{I[j]+1}\in\{0,1\}^d,
  \end{aligned}
\end{eqnarray*}
where $i,j\in \{B[0],B[1],...,B[n-1]\}$, 
let us define an $(r+d)$-qubit unitary operator $U_{x}$ ($x\in\{B[0],B[1],...,B[n-1]\}$) 
as follows, 
\begin{eqnarray*}
U_{x}
\equiv
\Big(
X^{x\oplus 1}\otimes I^{\otimes d}
\Big)
\cdot
\Lambda_{r}(X^{D[x]})
\cdot
\Big(
X^{x\oplus 1}\otimes I^{\otimes d}
\Big),
\end{eqnarray*}
where $\Lambda_{r}(X^{D[x]})$ is defined in Eq.~(\ref{eq:controlledgate}). 
Then it is clear that the following equation holds
\begin{eqnarray*}
\begin{aligned}
U_{x}
\Big(
|i\rangle \otimes |0^d\rangle
\Big)
=
  \left\{
    \begin{array}{l}
      |i\rangle \otimes |D[i]\rangle
      \ \ ({\rm{if}}\  i=x ), \\
      |i\rangle \otimes  |0^d\rangle
      \ \ ({\rm{otherwise}}),
    \end{array}
  \right.
\end{aligned}
\end{eqnarray*}
for any $r$-bit string $i$.
We also define $V_x$ ($x\in \{B[0],B[1],...,B[n-1]\}$) as
\begin{eqnarray*}
V_{x}
\equiv
\Big(
X^{x\oplus 1}\otimes I^{\otimes d}
\Big)
\cdot
\Lambda_{r}(X^{D'[x]})
\cdot
\Big(
X^{x\oplus 1}\otimes I^{\otimes d}
\Big),
\end{eqnarray*}
which encodes $D'[x]$ to a quantum state in the same way. 

Then it is possible to construct a unitary operation 
which corresponds to the QRAM operation of the above proof as
\begin{eqnarray}
&&(\prod_{x\in\{B[0],..,B[n-1]\}}U_xV_x)
\Big(\frac{1}{2^r}
\sum_{i,j\in\{0,1\}^r}|i\rangle_1\otimes |j\rangle_2
\nonumber\\
&&\ \ \  
\otimes |0^d\rangle_5 \otimes |0^d\rangle_6 
\Big)\nonumber\\
&&=
\frac{1}{2^r}
\sum_{i,j\in\{0,1\}^r}|i\rangle_1\otimes |j\rangle_2
\otimes |D[i]\rangle_5 \otimes |D'[j]\rangle_6.
\label{eq:OVunitary}
\end{eqnarray}
(We have omitted some registers for simplicity.) 

We consider a quantum circuit which just replaces the QRAM operation 
of the above proof with the unitary operation 
of Eq.~(\ref{eq:OVunitary}).
There is no need of additional ancilla qubit 
since the ancilla qubit for the generalized TOFFOLI gates 
can be used in common with that 
of the other steps of quantum computing. 
Thus the number of qubits used in this quantum computing 
is $N= 3d+3r+4$. 

The unitary operation of Eq.~(\ref{eq:OVunitary}) uses $\mathcal{O}(N^2 2^{\frac{N}{3(c+1)}})$ quantum gates. 
The reason is as follows: 
first, 
it is clear that this step uses $4nr$ $X$-gates at most. 
Next, 
$\Lambda_{r}(X^{D[x]})$ (and also $\Lambda_{r}(X^{D'[x]})$)
can be decomposed into $d$-number of generalized TOFFOLI gates at most and 
each $r$-qubit generalized TOFFOLI gate is composed of 
$8(r-3)$ TOFFOLI gates.
Therefore, $16nd(r-3)$ TOFFOLI gates are needed since we use $\Lambda_{r}(X^{D[x]})$ and $\Lambda_{r}(X^{D'[x]})$ $n$-times. 
Hence in total, the number of quantum gates used 
in this step is $\mathcal{O}(ndr)$ 
and
$\mathcal{O}(ndr)=\mathcal{O}(N^2 2^{\frac{N}{3(c+1)}})$
as it is seen from the following inequality:
\begin{eqnarray*}
  N=3d+3r+4
   &&\geq 3c\log_2{n}+3\log_2{n}+4\\
   &&> 3(c+1)\log_2{n}\\
\Leftrightarrow
  n&&<2^{\frac{N}{3(c+1)}},
\end{eqnarray*}
where we used Eq.~(\ref{eq:rofOV}).

Then, the size of this quantum computing is 
$\mathcal{O}(N^2 2^{\frac{N}{3(c+1)}})$ 
since the number of quantum gates used 
in the non-QRAM unitary operation is dominant as it 
is seen from table~\ref{tab:sizeOV}. 
The acceptance probability can also be defined to satisfy 
$p_{acc}={gap^2}/{2^{5r+3d+1}}$ 
in the same way. 
Thus, 
by applying the same argument as the above proof, 
this quantum computing cannot be exactly calculated 
in time $T\equiv2^{\frac{(2-\delta)(N-7)}{3(c+1)}}$ 
under Conjecture~\ref{conjecture:OV} 
and cannot be classically sampled within a multiplicative error 
$\epsilon<1$ in time $T$ under Conjecture~\ref{conjecture:OVgapNTIME}.
Hence Theorem \ref{theorem:ssOV} and \ref{theorem:wsOV} 
has been shown. 
\fbox

\section{3-SUM}
\label{sec:3SUM}
In this section, 
we show fine-grained quantum supremacy 
in terms of the qubit scaling 
based on 3-SUM and its variant. 
Let us introduce the following two conjectures:

\begin{conjecture}[3-SUM]
\label{conjecture:3SUM}
Given a set $S\subset\{-n^{3+\eta},...,n^{3+\eta}\}$ of size $n$, deciding
$s>0$ or $s=0$  
cannot be done
in time
$n^{2-\delta}$ 
for any $\eta,\delta>0$.
Here, 
\begin{eqnarray*}
s\equiv|\{(a,b,c)\in S\times S\times S~|~a+b+c=0\}|.
\end{eqnarray*}
\end{conjecture}

\begin{conjecture}
\label{conjecture:3SUMgapNTIME}
Given a set $S\subset\{-n^{3+\eta},...,n^{3+\eta}\}$ of size $n$, deciding
$gap\neq0$ or $gap=0$ cannot be done in 
non-deterministic time $n^{2-\delta}$ 
for any $\eta,\delta>0$. 
Here, 
\begin{eqnarray*}
gap\equiv|\{(a,&&b,c)\in S\times S\times S~|~a+b+c=0\}|\\
&&-|\{(a,b,c)\in S\times S\times S~|~a+b+c\neq0\}|.
\end{eqnarray*}
\end{conjecture}

Thinking of the QRAM model quantum computing, 
we can show the following results based on these two conjectures: 

\begin{theorem}[Strong simulation with QRAM]
\label{theorem:ss3SUMqram}
Assume that Conjecture~\ref{conjecture:3SUM} is true. Then
for any $\eta,\delta>0$, there exists an $N$-qubit 
and $\mathcal{O}(N)$-size 
quantum circuit with access to the QRAM 
whose acceptance probability cannot be
classically exactly calculated in 
$2^{\frac{(2-\delta)(N-18)}{13+3\eta}}$ time.
\end{theorem}

\begin{theorem}[Weak simulation with QRAM]
\label{theorem:ws3SUMqram}
Assume that Conjecture~\ref{conjecture:3SUMgapNTIME} is true. Then
for any $\eta,\delta>0$, there exists an $N$-qubit 
and $\mathcal{O}(N)$-size
quantum circuit with access to the QRAM 
whose acceptance probability cannot be classically
sampled within a multiplicative error $\epsilon<1$ in time
$2^{\frac{(2-\delta)(N-18)}{13+3\eta}}$.
\end{theorem}

By constructing a specific unitary operation 
corresponding to the QRAM operation, 
we can show the following results based on the above two conjectures: 

\begin{theorem}[Strong simulation]
\label{theorem:ss3SUM}
Assume that Conjecture~\ref{conjecture:3SUM} is true. Then
for any $\eta,\delta>0$, there exists an $N$-qubit 
and $\mathcal{O}(N^22^{\frac{N}{13+3\eta}})$-size quantum circuit 
whose acceptance probability cannot be
classically exactly calculated in 
$2^{\frac{(2-\delta)(N-18)}{13+3\eta}}$ time.
\end{theorem}

\begin{theorem}[Weak simulation]
\label{theorem:ws3SUM}
Assume that Conjecture~\ref{conjecture:3SUMgapNTIME} is true. Then
for any $\eta,\delta>0$, there exists an $N$-qubit 
and $\mathcal{O}(N^22^{\frac{N}{13+3\eta}})$-size quantum circuit 
whose acceptance probability cannot be classically
sampled within a multiplicative error $\epsilon<1$ in time
$2^{\frac{(2-\delta)(N-18)}{13+3\eta}}$.
\end{theorem}

{\it Proof of Theorem \ref{theorem:ss3SUMqram} and \ref{theorem:ws3SUMqram}}.
For a given set $S=\{e_1,...,e_n\}\subset\{-n^{3+\eta},...,n^{3+\eta}\}$ 
of size $n$, 
let us define the set $S'$ by 
\begin{eqnarray*}
S'\equiv\{e_1',e_2',...,e_n'\},
\end{eqnarray*}
where $e_i'\equiv e_i+n^{3+\eta}$ for all $i=1,2,...,n$.
Then, all elements of $S'$ are non-negative integers,
and $e_i+e_j+e_k=0$ if and only if 
$e_i'+e_j'+e_k'=3n^{3+\eta}$.
Let $r$ be the smallest integer such that $n\leq 2^r$ 
and $d$ be the smallest integer such that $2n^{3+\eta}\leq 2^d-1$, i.e., 
\begin{eqnarray}
\label{eq:rof3SUM}
\begin{aligned}
2^{r-1}<n\leq 2^r\hspace{1.2cm} \\
\Leftrightarrow
\log_2{n}\leq r<\log_2{n} +1,\\
\end{aligned}
\end{eqnarray}
and
\begin{eqnarray}
\label{eq:dof3SUM}
\begin{aligned}
2^{d-1}\leq 2n^{3+\eta}<2^d\hspace{1.55cm}\\
\Leftrightarrow
(3+\eta)\log_2{n}+1<d\leq (3+\eta)\log_2{n}+2.
\end{aligned}
\end{eqnarray}
Now we assume that we can use the QRAM 
which stores the data as 
\begin{eqnarray*}
D[i]=B[e_{I[i]+1}']\in \{0,1\}^d 
\end{eqnarray*}
for $i\in\{B[0],B[1],...,B[n-1]\}$. 
For such $i$ that satisfies $I[i]> n-1$, we assume $D[i]=0^d$. 

Let us consider the following quantum computing:
\begin{itemize}
\item[1.]
Generate
\begin{eqnarray*}
&&\frac{1}{\sqrt{2^{3r}}}
\sum_{i,j,k\in\{0,1\}^r}
|i\rangle_1\otimes
|j\rangle_2\otimes
|k\rangle_3\otimes
|B[n-1]\rangle_4 \otimes
|000\rangle_5 \\
&&\otimes |0^d\rangle_6 \otimes 
|0^{d+1}\rangle_7 \otimes |0^{d+2}\rangle_8 \otimes
|0\rangle_9.
\end{eqnarray*}
\item[2.]
Apply the quantum circuit $C$ of Eq.(\ref{eq:C})
which can compare two binary integers, 
between the 1st-4th, 2nd-4th and 3rd-4th registers, and 
flip the qubits of the 5th register according to their results, respectively:
\begin{eqnarray*}
&&\frac{1}{\sqrt{2^{3r}}}
\sum_{i,j,k\in\{0,1\}^r}
|i\rangle_1\otimes
|j\rangle_2\otimes
|k\rangle_3\otimes
|B[n-1]\rangle_4 \\
&&\otimes
|\chi(I[i]-n+1),\chi(I[j]-n+1),\chi(I[k]-n+1)\rangle_5 \\
&&\otimes
|0^d\rangle_6 \otimes |0^{d+1}\rangle_7 
\otimes |0^{d+2}\rangle_8 \otimes
|0\rangle_9,
\end{eqnarray*}
where $\chi(x)$ is defined in Eq.~(\ref{eq:chi}). 
Note that $|\chi(I[i]-n),\chi(I[j]-n),\chi(I[k]-n)\rangle$ is 
$|000\rangle$ 
if and only if $I[i]\leq n-1$, $I[j]\leq n-1$ and $I[k]\leq n-1$. 
\item[3.]
Apply the QRAM operation between the 1st-6th, 2nd-7th and 3rd-8th registers:
\begin{eqnarray*}
&&\frac{1}{\sqrt{2^{3r}}}
\sum_{i,j,k\in\{0,1\}^r}
|i\rangle_1\otimes
|j\rangle_2\otimes
|k\rangle_3\otimes
|B[n-1]\rangle_4 \\
&&\otimes
|\chi(I[i]-n+1),\chi(I[j]-n+1),\chi(I[k]-n+1)\rangle_5\\
&&\otimes 
|D[i]\rangle_6 \otimes |D[j],0\rangle_7 
\otimes |D[k],0,0\rangle_8 \otimes
|0\rangle_9.
\end{eqnarray*}
\item[4.]
Apply the addition circuit $A$ of Eq.~(\ref{eq:addition}) between the 6th and 7th
registers:
\begin{eqnarray*}
&&\frac{1}{\sqrt{2^{3r}}}
\sum_{i,j,k\in\{0,1\}^r}
|i\rangle_1\otimes
|j\rangle_2\otimes
|k\rangle_3\otimes
|B[n-1]\rangle_4 \\
&&\otimes
|\chi(I[i]-n+1),\chi(I[j]-n+1),\chi(I[k]-n+1)\rangle_5 \\
&&\otimes
|D[i]\rangle_6 \otimes |D[i]+D[j]\rangle_7 
\otimes |D[k],0,0\rangle_8 \otimes
|0\rangle_9,
\end{eqnarray*}
where $D[i]+D[j]$ is used in the meaning of 
$B[I[D[i]]+I[D[j]]]$. 
\item[5.]
Apply the addition circuit $A$ between the 7th and 8th
registers:
\begin{eqnarray*}
&&\frac{1}{\sqrt{2^{3r}}}
\sum_{i,j,k\in\{0,1\}^r}
|i\rangle_1\otimes
|j\rangle_2\otimes
|k\rangle_3\otimes
|B[n-1]\rangle_4 \\
&&\otimes
|\chi(I[i]-n+1),\chi(I[j]-n+1),\chi(I[k]-n+1)\rangle_5 
\otimes
|D[i]\rangle_6 \\
&&\otimes |D[i]+D[j]\rangle_7 \otimes |D[i]+D[j]+D[k]\rangle_8 \otimes
|0\rangle_9.
\end{eqnarray*} 

\item[6.]
Flip the last register if
the 8th register encodes $3n^{3+\eta}$, 
by applying 
\begin{eqnarray*}
(X^{B[3n^{3+\eta}]\otimes I})\cdot(\Lambda_{(d+2)}{(X)})\cdot(X^{B[3n^{3+\eta}]\otimes I})
\end{eqnarray*}
between the 8th and 9th registers:
\begin{eqnarray*}
&&\frac{1}{\sqrt{2^{3r}}}
\sum_{i,j,k\in\{0,1\}^r}
|i\rangle_1\otimes
|j\rangle_2\otimes
|k\rangle_3\otimes
|B[n-1]\rangle_4 \\
&&\otimes
|\chi(I[i]-n+1),\chi(I[j]-n+1),\chi(I[k]-n+)\rangle_5\\ &&\otimes
|D[i]\rangle_6 
\otimes |D[i]+D[j]\rangle_7 \otimes |D[i]+D[j]+D[k]\rangle_8 \\ &&\otimes 
|\delta_{D[i]+D[j]+D[k],3n^{3+\eta}}\rangle_9.
\end{eqnarray*}

\item[7.]
Apply $Z$ gate to the last qubit and finally get
\begin{eqnarray*}
&&\frac{1}{\sqrt{2^{3r}}}
\sum_{i,j,k\in\{0,1\}^r}
(-1)^{\delta_{D[i]+D[j]+D[k],3n^{3+\eta}}}
|i\rangle_1\otimes
|j\rangle_2\otimes
|k\rangle_3\\
&&\otimes
|B[n-1]\rangle_4 \\
&&\otimes
|\chi(I[i]-n+1),\chi(I[j]-n+1),\chi(I[k]-n+1)\rangle_5 \\
&&\otimes|D[i]\rangle_6 \otimes |D[i]+D[j]\rangle_7 
\otimes |D[i]+D[j]+D[k]\rangle_8 \\
&&\otimes
|\delta_{D[i]+D[j]+D[k],3n^{3+\eta}}\rangle_9\equiv |\Phi\rangle. 
\end{eqnarray*}

\item[8.]
Measure qubits of the 5th register of $|\Phi\rangle$ 
in the $Z$ basis and measure all the other qubits of $|\Phi\rangle$ 
in the $X$ basis. 
If all results are $0$, 
then accept. 
Then, 
the acceptance probability is 
\begin{eqnarray}
\label{eq:pacc3SUM}
\begin{aligned}
p_{acc}
&\equiv
|\langle +^{4r}000+^d+^{d+1}+^{d+2}+|\Phi\rangle|^2&\\
&=
\frac{gap^2}{2^{7r+3d+4}}.&
\end{aligned}
\end{eqnarray}

\end{itemize}

This quantum computing needs $4r+3d+8$ qubits, because of the following
reasons: first, we used $4r+3d+7$ qubits as an initial state.
Second,
each of the quantum circuit $C$, $A$ and the generalized TOFFOLI gate used in the above quantum computing needs a single ancilla qubit, which can be used in common.  
Hence $4r+3d+8\equiv N$ qubits are needed in total. Then the following inequality holds using Eq.~(\ref{eq:rof3SUM}) and (\ref{eq:dof3SUM}):
\begin{eqnarray*}
N=4r+3d+8<(13+3\eta)\log_2{n}+18.
\end{eqnarray*}

\begin{table}[htb]
  \caption{The number of quantum gates 
  used at most in each step of the quantum computation
  of 3-SUM.}
  \begin{tabular}{|c|c|r|} \hline
    step & gate &number \\ \hline \hline
    1.   & $H$-gate  & $3r$     \\ 
         & $X$-gate  & $r$     \\ \hline
    2.   & $X$-gate  & $6r+9$     \\ 
         & $CX$-gate  & $12r+3$     \\
         & TOFFOLI  & $6r$     \\ \hline
    3.   & QRAM  &  $3$   \\ \hline
    4.   & $CX$-gate  & $4d+1$     \\
         & TOFFOLI  & $2d$     \\ \hline   
    5.   & $CX$-gate  & $4d+5$     \\
         & TOFFOLI  & $2d+2$     \\ \hline 
    6.   & $X$-gate  & $2d+4$     \\ 
         & TOFFOLI  & $8(d-1)$     \\ \hline            
    7.   & $Z$-gate      & 1 \\ \hline
   Non-QRAM 
         & $X$-gate      & $6nr$ \\ 
   unitary operation      & TOFFOLI       & $24nd(r-3)$ \\ \hline
  \end{tabular}
  \label{tab:size3SUM}
\end{table}

We summarize the number of quantum gates used 
at most in each step of quantum computation in 
table~\ref{tab:size3SUM}.
As it can be seen from this table, 
this quantum computing is of $\mathcal{O}(N)$ size.

Let us define $T$ as
\begin{eqnarray*}
T\equiv2^{\frac{(2-\delta)(N-18)}{13+3\eta}}
<n^{2-\delta}.
\end{eqnarray*}
Assume that $p_{acc}$ of Eq.~(\ref{eq:pacc3SUM}) is classically exactly calculated in time $T$. Then,
$|\{(a,b,c)\in S\times S\times S~|~a+b+c=0\}|=(gap+n^3)/2>0$ or $=0$
can be decided in time $n^{2-\delta}$, which contradicts to 
Conjecture~\ref{conjecture:3SUM}.
Hence Theorem~\ref{theorem:ss3SUMqram} has been shown.
Next assume that $p_{acc}$ is classically sampled within a multiplicative
error $\epsilon<1$ in time $T$.
Then,
$gap\neq0$ or $=0$ can be decided in non-deterministic
time $n^{2-\delta}$, which
contradicts to
Conjecture~\ref{conjecture:3SUMgapNTIME}.
Hence Theorem~\ref{theorem:ws3SUMqram} has been shown.\fbox

\vspace{0.3cm}
{\it Proof of Theorem \ref{theorem:ss3SUM} and \ref{theorem:ws3SUM}}.
Let us define an $(r+d)$-qubit unitary operator $U_{x}$ 
($x\in\{B[0],B[1],...,B[n-1]\}$) 
as follows, 
\begin{eqnarray*}
\begin{aligned}
U_{x}
\equiv
\Big(
X^{x\oplus 1}\otimes I^{\otimes d}
\Big)
\cdot
\Lambda_{r}(X^{D[x]})
\cdot
\Big(
X^{x\oplus 1}\otimes I^{\otimes d}
\Big),
\end{aligned}
\end{eqnarray*}
where $\Lambda_{r}(X^{D[x]})$ is defined in Eq.~(\ref{eq:controlledgate}). 
Then it is clear that
\begin{eqnarray*}
\begin{aligned}
U_{x}
\Big(
|i\rangle \otimes |0\rangle^{\otimes d}
\Big)
=
  \left\{
    \begin{array}{l}
      |i\rangle \otimes  |D[i]\rangle
      \ \ \ ({\rm{if}}\  x=i ), \\
      |i\rangle \otimes |0\rangle^{\otimes d}
      \ \ \ ({\rm{otherwise}}),
    \end{array}
  \right.
\end{aligned}
\end{eqnarray*}
for any $r$-bit string $i$. 
We can realize a step which corresponds to the QRAM operation 
of the above proof 
by applying 
$\Big(\prod_{x\in\{B[0],B[1],...,B[n-1]\}}U_{x}\Big)$ between 
the 1st-6th, 2nd-7th and 3rd-8th registers of 
the quantum state of step 2. 
This step needs $\mathcal{O}(nrd)$ quantum gates 
because each $\Lambda_{r}(X^{D[x]})$ 
in $U_x$ is composed of at most 
$d$-number of $r$-controlled generalized TOFFOLI gates 
and we use $U_x$ $n$ times 
while the number of $X$-gate used in this step is $\mathcal{O}(nr)$.

We consider a quantum circuit which just replaces the QRAM operation of the above proof with this unitary operation.
There is no need of additional ancilla qubit 
for this replacement 
because the ancilla qubit for the generalized TOFFOLI gates 
can be used in common with the ancilla qubit used 
in other steps of quantum computing. 
Therefore, 
the number of qubits used in this quantum computing is $N=4r+3d+8$. 
As it can be seen from table~\ref{tab:size3SUM}, the quantum computing without the QRAM has $\mathcal{O}(nrd)$ size, 
and 
$\mathcal{O}(nrd)=\mathcal{O}(N^2 2^{\frac{N}{13+3\eta}})$ because it follows from 
Eq.~(\ref{eq:rof3SUM}) and Eq.~(\ref{eq:dof3SUM}) that
\begin{eqnarray*}
N=4r+3d+8 
&&> 4\log_2{n}+3(3+\eta)\log_2{n}+11\\
&&> (13+3\eta)\log_2{n}\\
\Leftrightarrow
n&&<2^{\frac{N}{13+3\eta}}.
\end{eqnarray*}
Hence by applying the same argument with the above proof, 
Theorem \ref{theorem:ss3SUM} and \ref{theorem:ws3SUM} have been shown.
\fbox

\section{Negative Weight Triangle}
\label{sec:nwt}
In this section, 
we show fine-grained quantum supremacy 
in terms of the qubit scaling 
based on Negative Weight Triangle and its variant. 
Let us introduce the following two conjectures:

\begin{conjecture}[Negative Weight Triangle]
\label{conjecture:NWT}
Given an edge-weighted $n$-vertex graph $G=(V,E)$ 
with integer weights from $\lbrace -M,...,M \rbrace$,
where $M$ is a certain integer,
deciding  whether $s>0$ or $s=0$ needs $n^{3-\delta}$ time for any $\delta > 0$.
Here, 
\begin{eqnarray*}
s\equiv \big|\lbrace (i,j,k) \in V^3|(i,j,k)\text{ is good}\rbrace\big|,
\end{eqnarray*} 
where we say $(i,j,k)$ is good if it is triangle and 
\begin{equation*}
W(e_{i,j})+W(e_{j,k})+W(e_{k,i})<0,
\end{equation*}
where $e_{i,j}$ is the edge between vertices $i$ and $j$, 
and $W(e_{i,j})$ is the weight of it.
Note that $W(e_{i,j})=0$ means that 
the edge $e_{i,j}$ has weight $0$, 
which is different from no-edge.
\end{conjecture}

\begin{conjecture}
\label{conjecture:NWTva}
Given an edge-weighted $n$-vertex graph $G=(V,E)$ with integer weights 
from $\lbrace -M,...,M\rbrace$, 
where $M$ is a certain integer,
deciding whether $gap\neq0$ or $gap=0$ needs non-deterministic 
$n^{3-\delta}$ time for any $\delta > 0$.
Here, 
\begin{eqnarray*}
gap \equiv
\bigl| &&\lbrace (i,j,k) \in V^3 |(i,j,k) \text{ is good}  \rbrace \bigr|\\
&&-\bigl|\lbrace (i,j,k) \in V^3 |(i,j,k)\text{ is not good}\rbrace \bigr|.
\end{eqnarray*}
\end{conjecture}

Thinking of the QRAM model quantum computing, 
we can show the following two results 
based on the above two conjectures:

\begin{theorem}[Strong Simulation with QRAM]
\label{theorem:ssNWTqram}
Assume that Conjecture~\ref{conjecture:NWT} is true.
Then,
for any $\delta > 0$, there is an $M$ such that there exists an $N$-qubit
and $\mathcal{O}(N)$-size quantum circuit 
with access to the QRAM 
whose acceptance probability cannot be classically exactly calculated 
in time $2^{\frac{3-\delta}{4}(N-4\log_2(2M+1)-22)}$.
\end{theorem}

\begin{theorem}[Weak Simulation with QRAM]
\label{theorem:wsNWTqram}
Assume that Conjecture~\ref{conjecture:NWTva} is true. 
Then,
for any $\delta > 0$, 
there is an $M$ such that 
there exists an $N$-qubit 
and $\mathcal{O}(N)$-size quantum circuit 
with access to the QRAM 
whose acceptance probability cannot be classically sampled 
within a multiplicative error $\epsilon < 1$ in time $2^{\frac{3-\delta}{4}(N-4\log_2(2M+1)-22)}$.
\end{theorem}

By constructing a specific unitary operation 
corresponding to the QRAM process, 
we can show the following two results 
based on the above two conjectures:

\begin{theorem}[Strong Simulation]
\label{theorem:ssNWT}
Assume that Conjecture~\ref{conjecture:NWT} is true.
Then,
for any $\delta > 0$, there is an $M$ such that there exists 
an $N$-qubit and $\mathcal{O}(2^{\frac{N}{2}} N^2)$-size quantum circuit 
whose acceptance probability cannot be classically exactly calculated 
in time $2^{\frac{3-\delta}{4}(N-4\log_2(2M+1)-22)}$.
\end{theorem}

\begin{theorem}[Weak Simulation]
\label{theorem:wsNWT}
Assume that Conjecture~\ref{conjecture:NWTva} is true.
Then, 
for any $\delta > 0$, 
there is an $M$ such that 
there exists 
an $N$-qubit and $\mathcal{O}(2^{\frac{N}{2}} N^2)$-size quantum circuit 
whose acceptance probability cannot be classically sampled 
within a multiplicative error $\epsilon < 1$ in time $2^{\frac{3-\delta}{4}(N-4\log_2(2M+1)-22)}$.
\end{theorem}

{\it Proof of theorem \ref{theorem:ssNWTqram} and \ref{theorem:wsNWTqram}}.
For a given edge-weighted $n$-vertex graph $G=(V,E)$ 
with integer weights from $\lbrace -M,...,M \rbrace$, 
let us define two integers $r$ and $s$ to satisfy
\begin{eqnarray}
\label{eq:rofNWT}
2^{r-1}<&n&\leq 2^r\nonumber\\
\Leftrightarrow
\log_2{n}\leq &r& < \log_2{n}+1
\end{eqnarray}
and
\begin{eqnarray}
\label{eq:dofNWT}
2^{d-1}\leq 2M+1< 2^d\nonumber\hspace{2cm}\\
\Leftrightarrow
\log_2{(2M+1)}< d \leq \log_2{(2M+1)}+1.
\end{eqnarray}
We can think of a corresponding adjacency matrix $A_{ij}$ ($i,j \in\{0,1\}^r$) 
which is defined as 
\begin{eqnarray*}
&{}&A_{ij}\equiv \\
&{}& \left \{
\begin{array}{cl}
&W(e_{I[i]+1,I[j]+1}) \ \ \ 
\begin{aligned}
&{\rm{if\ vertices}}\ I[i]+1\ {\rm{and}}\ I[j]+1\ \\
&\ \ \ {\rm{have\ an\ edge}},
\end{aligned}
\\
&M+1\hspace{1cm}
\begin{aligned}
&{\rm{if\ vertices}}\ I[i]+1\ {\rm{and}}\ I[j]+1\ \\
&\ \ {\rm{do\ not\ have\ an\ edge}},
\end{aligned}
\\
&M+1\hspace{1cm} {\rm{if}}\ i=j,\\
&M+1\hspace{1cm} {\rm{if}}\ I[i]\geq n \ {\rm{or}}\ I[j]\geq n.
\end{array}
\right.
\end{eqnarray*}
In order to restrict all matrix elements to be non-negative, 
we define matrix $W$ from $A$ as $W_{ij}\equiv A_{ij} + M \in \{0,1,..,2M+1\}$ 
for all $i,j\in\{0,1\}^r$:
\begin{eqnarray*}
    &{}&W_{ij}\equiv \\
    &{}& \left \{
    \begin{array}{cl}
    &W(e_{I[i]+1,I[j]+1})+M \ \ \ 
    \begin{aligned}
      &{\rm{if\ vertices}}\ I[i]+1\ {\rm{and}}\ I[j]+1\ \\
      &\ \ {\rm{have\ an\ edge}},
    \end{aligned}\\
    &2M+1\ \ \ \  \ 
    \begin{aligned}
      &{\rm{if\ vertices}}\ I[i]+1\ {\rm{and}}\ I[j]+1\ \\
      &\ \ {\rm{do\ not\ have\ an\ edge}},
      \end{aligned}\\
    &2M+1\ \ \ \  \ {\rm{if}}\ i=j, \\
    &2M+1\ \ \ \  \ {\rm{if}}\ I[i]\geq n \ {\rm{or}}\ I[j]\geq n.
    \end{array}
    \right.
\end{eqnarray*}

We assume that we can access to the QRAM which returns the data 
by inputting two binary strings as 
\begin{eqnarray}
\label{eq:nwtQRAM}
\sum_{x,y\in\{0,1\}^r} |x\rangle\otimes |y\rangle\otimes|0^d\rangle 
\rightarrow
\sum_{x,y} |x\rangle \otimes|y \rangle\otimes|B[W_{xy}]\rangle. \ \ \ \ \ \ 
\end{eqnarray}
We define an $(d+1)$-qubit unitary gate $V$ as 
\begin{eqnarray}
V
\equiv
\Big(X^{B[2M+1]\oplus 1}\otimes I\Big)
\cdot
\Lambda_{d}(X)
\cdot
\Big(X^{B[2M+1]\oplus 1}\otimes I\Big),\ \ \ \ \ \ 
\label{eq:v}
\end{eqnarray}
where $\Lambda_{d}(X)$ is the $d$-controlled $X$ gate.
Then, 
it is clear that 
\begin{eqnarray*}
\begin{aligned}
V
\Big(
|w\rangle \otimes |0\rangle
\Big)
=
  \left\{
    \begin{array}{l}
      |w\rangle \otimes |1\rangle
      \ \ ({\rm{if}}\  w=B[2M+1]), \\
      |w\rangle \otimes |0\rangle
      \ \ ({\rm{otherwise}}),
    \end{array}
  \right.
\end{aligned}
\end{eqnarray*}
for any $d$-bit string $w$. 

Let us consider the following 
quantum computing:

\begin{itemize}
\item[1.]
First, we generate the following $(4r+3)$-qubit quantum state, 
\begin{eqnarray*}
\begin{aligned}
|&\varphi_0\rangle
\equiv\\
&\frac{1}{\sqrt{2^{3r}}}
\sum_{x,y,z\in\lbrace0,1\rbrace^r}
|x\rangle_1 \otimes|y\rangle_2 \otimes|z\rangle_3 \otimes
|B[n-1]\rangle_4\otimes
|000\rangle_5.
\end{aligned}
\end{eqnarray*}

\item[2.]
Next, 
we apply the quantum circuit $C$ of Eq.~(\ref{eq:C}) between the 1st-4th, 2nd-4th and 3rd-4th registers, 
and flip the qubits of the 5th register according to their results, respectively:
\begin{eqnarray*}
\begin{aligned}
|\varphi_1\rangle
&=
\frac{1}{\sqrt{2^{3r}}}
\sum_{x,y,z\in\lbrace0,1\rbrace^r}
|x\rangle_1\otimes|y\rangle_2\otimes|z\rangle_3\otimes
|B[n-1]\rangle_4\\
\otimes &|\chi (I[x]-n+1),\chi(I[y]-n+1),\chi(I[z]-n+1)\rangle_5
\\
&\equiv \sum_{x,y,z\in\lbrace0,1\rbrace^r} |h(x,y,z)\rangle_{1\sim 5}.
\end{aligned}
\end{eqnarray*} 
We have defined $|h(x,y,z)\rangle_{1\sim 5}$ 
to simplify the notation. 
Note that 
$|\chi (I[x]-n+1), \chi(I[y]-n+1), \chi(I[z]-n+1)\rangle$ 
is $|000\rangle$ if and only if $I[x]\leq n-1$, $I[y]\leq n-1$ and $I[z]\leq n-1$.

\item[3.]
Next, 
we add $|0^d\rangle\otimes|0^{d+1}\rangle\otimes|0^{d+2}\rangle\otimes
|0^3\rangle\otimes |0^{d+2}\rangle\otimes
|0\rangle \otimes|0\rangle$ to $|\varphi_1\rangle$ and get
\begin{eqnarray*}
\begin{aligned}
|\varphi_2\rangle
=
&\sum_{x,y,z\in\lbrace0,1\rbrace^r} |h(x,y,z)\rangle_{1\sim 5}
\otimes
|0^d\rangle_6
\otimes
|0^{d+1}\rangle_7
\otimes
|0^{d+2}\rangle_8\\
&\ \otimes 
|0^3\rangle_9\otimes |0^{d+2}\rangle_{10}\otimes
|0\rangle_{11} \otimes|0\rangle_{12}.
\end{aligned}
\end{eqnarray*}
Now we use the QRAM of Eq.~(\ref{eq:nwtQRAM}) between the
1st-2nd-6th, 2nd-3rd-7th and 1st-3rd-8th registers of $|\varphi_2\rangle$. 
Then we get
\begin{eqnarray*}
\begin{aligned}
|\varphi_3\rangle
=
&\sum_{x,y,z\in\lbrace0,1\rbrace^r} |h(x,y,z)\rangle_{1\sim 5}
\otimes |B[W_{xy}]\rangle_6\otimes |B[W_{yz}],0\rangle_7 \\
&\otimes |B[W_{xz}],0,0\rangle_8
\otimes
|0^3\rangle_9\otimes |0^{d+2}\rangle_{10}\otimes
|0\rangle_{11} \otimes|0\rangle_{12}.
\end{aligned}
\end{eqnarray*}
\item[4.]
We use the ($d+1$)-qubit operator $V$ defined in Eq.~(\ref{eq:v}). 
We apply $V$ between 
the 6th-$9_1$th, 
7th-$9_2$th and 8th-$9_3$th registers of $|\varphi_3\rangle$, 
where $9_i$ means the $i$th qubit of the 9th register. 
Then we get
\begin{eqnarray*}
\begin{aligned}
|\varphi_4\rangle
=
&\Big(V_{6{\mathchar`-}9_1}\Big)
\Big(V_{7{\mathchar`-}9_2}\Big)
\Big(V_{8{\mathchar`-}9_3}\Big)
|\varphi_3\rangle\\
=
& \sum_{x,y,z\in\lbrace0,1\rbrace^r} |h(x,y,z)\rangle_{1\sim 5}\otimes |B[W_{xy}]\rangle_6\otimes |B[W_{yz}],0\rangle_7 \\
&\otimes |B[W_{xz}],0,0\rangle_8
\otimes
|f(W_{xy}),f(W_{yz}),f(W_{zx})\rangle_9
\\
&\otimes
|0^{d+2}\rangle_{10}\otimes
|0\rangle_{11} \otimes|0\rangle_{12},
\end{aligned}
\end{eqnarray*}
where
\begin{eqnarray*}
 f(p)=
  \left\{
    \begin{array}{l}
      1 \ (p=2M+1), \\
      0 \ (\rm{otherwise}).
    \end{array}
  \right.
\end{eqnarray*}

\item[5.]
Apply the addition circuit $A$ of Eq.~(\ref{eq:addition}) between the 6th-7th registers of 
$|\varphi_4\rangle$,  
and apply $A$ again between the 7th-8th registers. 
Then we get
\begin{eqnarray*}
\begin{aligned}
|\varphi_5\rangle
=
&\sum_{x,y,z\in\lbrace0,1\rbrace^r} |h(x,y,z)\rangle_{1\sim 5}
\otimes |B[W_{xy}]\rangle_6 \\
&\otimes |B[W_{xy}+W_{yz}]\rangle_7 
\otimes |B[W_{xy}+W_{yz}+W_{xz}]\rangle_8\\
&\otimes
|f(W_{xy}), f(W_{yz}),f(W_{zx})\rangle_9\otimes
|0^{d+2}\rangle_{10}\\
&\otimes 
|0\rangle_{11} \otimes|0\rangle_{12}.
\end{aligned}
\end{eqnarray*}
Note that $B[W_{xy}+W_{yz}]$ and $B[W_{xy}+W_{yz}+W_{xz}]$ 
are represented in $d+1$ and $d+2$ bit strings, respectively. 

\item[6.] 
First we apply $X^{B[3M]}$ to the 10th register of $|\varphi_5\rangle$. 
After this, 
we apply the quantum circuit $C'$ of Eq.~(\ref{eq:Cprime}) 
between the 8th-10th registers 
and flip the qubit of the 11th register according to the result. 
Then we get
\begin{eqnarray*}
\begin{aligned}
|\varphi_6\rangle
=
&\sum_{x,y,z\in\lbrace0,1\rbrace^r} |h(x,y,z)\rangle_{1\sim 5}
\otimes |B[W_{xy}]\rangle_6 \\
&\otimes |B[W_{xy}+W_{yz}]\rangle_7 
\otimes |B[W_{xy}+W_{yz}+W_{xz}]\rangle_8\\
&\otimes
|f(W_{xy}), f(W_{yz}),f(W_{zx})\rangle_9\otimes
|B[3M]\rangle_{10}\\
&\otimes|\chi(3M-(W_{xy}+W_{yz}+W_{xz}))\oplus 1\rangle_{11} \otimes|0\rangle_{12}.
\end{aligned}
\end{eqnarray*}

\item[7.] 
Flip the last register 
if all of the qubits of the 9th and 11th registers of $|\varphi_6\rangle$ are $0$.
Then we get
\begin{eqnarray*}
\begin{aligned}
|\varphi_7\rangle
&=
\sum_{x,y,z\in\lbrace0,1\rbrace^r} |h(x,y,z)\rangle_{1\sim 5}
\otimes |B[W_{xy}]\rangle_6 \\
&\otimes |B[W_{xy}+W_{yz}]\rangle_7 
\otimes |B[W_{xy}+W_{yz}+W_{xz}]\rangle_8\\
&\otimes
|f(W_{xy}), f(W_{yz}),f(W_{zx})\rangle_9\otimes
|B[3M]\rangle_{10}\\
&\otimes|\chi(3M-(W_{xy}+W_{yz}+W_{xz}))\oplus 1\rangle_{11} \\
&\otimes|g(x,y,z)\rangle_{12},
\end{aligned}
\end{eqnarray*}
where 
\begin{eqnarray*}
&&g(x,y,z)\\=
&&\begin{cases}
    1 & 
    \begin{aligned}
       ({\rm if}\ W_{xy}\neq 2M+1 \ 
       \cap\ W_{yz}\neq 2M+1 
       \cap W_{zx}\\ \neq 2M+1 
       \cap\ W_{xy}+W_{yz}+W_{xz}<3M ),
    \end{aligned} \\
    0 & ({\rm otherwise}).
\end{cases}
\end{eqnarray*}

\item[8.]
Apply $Z$ gate to the last qubit of $|\varphi_7\rangle$ 
and finally get
\begin{eqnarray}
\label{eq:PhiNWT}
\begin{aligned}
\frac{1}{\sqrt{2^{3r}}} 
&\sum_{x,y,z\in\lbrace0,1\rbrace^r}
(-1)^{g(x,y,z)}
|x\rangle_1\otimes|y\rangle_2\otimes|z\rangle_3\otimes
|B[n-1]\rangle_4
\\
&\otimes|\chi (I[x]-n+1),\chi(I[y]-n+1),\chi(I[z]-n+1)\rangle_5
\\
&\otimes |B[W_{xy}]\rangle_6\otimes |B[W_{xy}+W_{yz}]\rangle_7 \\
&\otimes |B[W_{xy}+W_{yz}+W_{xz}]\rangle_8\\
&\otimes
|f(W_{xy}), f(W_{yz}),f(W_{zx})\rangle_9\otimes
|B[3M]\rangle_{10}\\
&\otimes|\chi(3M-(W_{xy}+W_{yz}+W_{xz}))\oplus 1\rangle_{11} 
\otimes|g(x,y,z)\rangle_{12}\\
&\ \ \ \equiv
|\Phi\rangle.
\end{aligned}
\end{eqnarray}

\item[9.]
Measure qubits of the 5th register of $|\Phi\rangle$ 
in the $Z$ basis
and 
measure all the other qubits of $|\Phi\rangle$ 
in the $X$ basis. 
If all results are $0$, 
then accept. 
Then, 
the acceptance probability is 
\begin{eqnarray}
\label{eq:paccNWT} 
\begin{aligned}
p_{acc}
\equiv
&
|\langle
+^{4r}0^3+^{4d+10}|\Phi
\rangle |^2\\
=
&\frac{gap^2}{2^{7r+4d+10}}.
\end{aligned}
\end{eqnarray} 
\end{itemize}

This quantum computing needs $4r+4d+14$ qubits, since
we prepared $4r+3$ qubits in the 1st step and we added $4d+10$ qubits in the 3rd step. 
We need an additional ancilla qubit which is used in common for the quantum circuits 
$A$, $C$, $C'$ and the generalized TOFFOLI gates. 
Hence $4r+4d+14 \equiv N$ qubits are needed in total. The following inequality holds using Eq.~(\ref{eq:rofNWT}) and Eq.~(\ref{eq:dofNWT}):
\begin{eqnarray*}
N=4r+4d+14<4\log_2{n}+4\log_2{(2M+1)}+22.
\end{eqnarray*}

\begin{table}[htb]
  \caption{The number of quantum gates 
  used at most in each step of the quantum computation
  of NWT.}
  \begin{tabular}{|c|c|r|} \hline
    step & gate       & number   \\ \hline \hline
    1.   & $H$-gate   & $3r$     \\ 
         & $X$-gate   & $r$       \\ \hline
    2. 
         & $X$-gate   & $6r+9$     \\ 
         & $CX$-gate  & $12r+3$    \\
         & TOFFOLI    & $6r$       \\ \hline
    3.   & QRAM       & $3$         \\ \hline
    4.   & $X$-gate   & $6d$        \\
         & TOFFOLI    & $24(d-3)$    \\ \hline   
    5.   & $CX$-gate  & $8d+6$       \\
         & TOFFOLI    & $4d+2$       \\ \hline 
    6.   & $X$-gate   & $3d+8$       \\ 
         & $CX$-gate  & $4d+9$        \\
         & TOFFOLI    & $2d+4$        \\ \hline          
    7.   & $X$-gate   & 8           \\ 
         & TOFFOLI    & 10          \\ \hline
    8.   & $Z$-gate   & 1            \\ \hline
   Non-QRAM 
         & $X$-gate      & $12r2^{2r}$ \\ 
   unitary operation      
         & TOFFOLI       & $24d(2r-3)2^{2r}$ \\ \hline
  \end{tabular}
  \label{tab:sizeNWT}
\end{table}

We summarize the number of quantum gates used in each step at most 
in table~\ref{tab:sizeNWT}. 
As it can be seen from this table, 
this quantum computing uses $\mathcal{O}(N)$ gates. 

Then, 
let us define $T$ by
\begin{eqnarray*}
T \equiv 2^{\frac{(3-\delta)}{4}(N-4\log_2(2M+1)-22)}
<
n^{3-\delta}.
\end{eqnarray*}
Assume that $p_{acc}$ of Eq.~(\ref{eq:paccNWT}) is classically exactly calculated in time $T$. 
Then, 
$s=(gap+n^3)/2>0$ or $s=0$ can be decided in time $n^{3-\delta}$, 
which contradicts to Conjecture ~\ref{conjecture:NWT}. 
Hence, Theorem ~\ref{theorem:ssNWTqram} has been shown. 
Next, 
assume that $p_{acc}$ can be classically sampled 
within a multiplicative error $\epsilon<1$ in time $T$.
Then, $gap \neq 0$ or $=0$ can be decided in 
non-deterministic time $n^{3-\delta}$, 
which contradicts to Conjecture~\ref{conjecture:NWTva}. 
Hence Theorem~\ref{theorem:wsNWTqram} has been shown.\fbox

\vspace{0.3cm}
{\it Proof of Theorem \ref{theorem:ssNWT} and \ref{theorem:wsNWT}}.
Let us define an $(2r+d)$-qubit unitary operator 
$U_{ij}$ ($i,j\in\{0,1\}^r$) 
as follows, 
\begin{eqnarray*}
\begin{aligned}
U_{ij}
\equiv
\Big(
&X^{B[i]\oplus 1}\otimes X^{B[j]\oplus 1}\otimes I^{\otimes d}
\Big)
\cdot
\Lambda_{2r}(X^{B[W_{ij}]})\\
&\cdot
\Big(
X^{B[i]\oplus 1}
\otimes
X^{B[j]\oplus 1}\otimes
I^{\otimes d}
\Big),
\label{eq:unitary}
\end{aligned}
\end{eqnarray*}
where $\Lambda_{2r}(X^{B[W_{ij}]})$ is 
defined in Eq.~(\ref{eq:controlledgate}). 
Then it is clear that the following equation holds
\begin{eqnarray*}
\begin{aligned}
U_{ij}
\Big(
|x\rangle &\otimes |y\rangle \otimes |0^d\rangle
\Big)\\
&=
  \left\{
    \begin{array}{l}
      |x\rangle \otimes |y\rangle \otimes |B[W_{ij}]\rangle
      \ \ ({\rm{if}}\  x=i \ {\rm{and}} \ y=j) \\
      |x\rangle \otimes |y\rangle \otimes |0^d\rangle
      \ \ \ \ \ \ \ \ ({\rm{otherwise}}),
    \end{array}
  \right.
\end{aligned}
\end{eqnarray*}
for any $r$-bit strings $x$ and $y$.
We can realize a unitary operation 
which corresponds to the QRAM operation  
of the above proof 
by applying $\Big(\prod_{i,j\in\{0,1\}^r}U_{ij}\Big)$ between the
1st-2nd-6th, 2nd-3rd-7th  and 1st-3rd-8th registers of $|\varphi_2\rangle$ as
\begin{eqnarray*}
\begin{aligned}
|\varphi_3\rangle
&=
\Big(\prod_{i,j\in\{0,1\}^r}(U_{ij})_{1{\mathchar`-}2{\mathchar`-}6}
(U_{ij})_{2{\mathchar`-}3{\mathchar`-}7}
(U_{ij})_{1{\mathchar`-}3{\mathchar`-}8}\Big)
\ |\varphi_2\rangle\\
&=
\sum_{x,y,z\in\lbrace0,1\rbrace^r} |h(x,y,z)\rangle_{1\sim 5}
\\
&\ \ \otimes |B[W_{xy}]\rangle_6\otimes |B[W_{yz}],0\rangle_7 \otimes |B[W_{xz}],0,0\rangle_8\\
&\ \ \otimes |0^3\rangle_9\otimes |0^{d+2}\rangle_{10}\otimes
|0\rangle_{11} \otimes|0\rangle_{12}.
\end{aligned}
\end{eqnarray*}
This unitary operation uses $\mathcal{O}(2^{2r} dr)$ 
quantum gates because each 
of the $\Lambda_{2r}(X^{B[Wij]})$ 
is composed of at most $d$-number of 
$2r$-qubit controlled generalized TOFFOLI gate 
and 
we use $U_{ij}$ $3(2^r)^2$ times. 
Therefore, the number of TOFFOLI gates used in this step is $\mathcal{O}(2^{2r} dr)$ 
while the number of $X$ gates used in this operation is $\mathcal{O}(2^{2r}r)$. 
Thus $\mathcal{O}(2^{2r} dr)$ size is required in this step.

We consider a quantum circuit which just replaces the QRAM operation of the above proof with this unitary operation.
There is no need of additional ancilla qubit 
for this replacement 
because the ancilla qubit for the generalized TOFFOLI gates 
can be used in common with that of the other steps. 
Therefore, this quantum computing uses $N=4r+4d+14$ qubits. 
The size of this quantum computing is $\mathcal{O}(2^{2r} dr)$ as it is seen from table~\ref{tab:sizeNWT}, and $\mathcal{O}(2^{2r} dr)=\mathcal{O}(2^{\frac{N}{2}} N^2)$
since $2r=\frac{N-4d-14}{2}<\frac{N}{2}$. 
Hence by applying the same argument with the above proof, 
Theorem \ref{theorem:ssNWTqram} and \ref{theorem:wsNWTqram} have been shown.
\fbox

\section{discussion}
In this paper, we have considered the worst-case hardness, but it
would be an interesting open problem to show fine-grained quantum
supremacy for the average case~\cite{averagehardness}.

The results of this paper can be reduced to those of several sub-universal models of quantum computing. 
First, we consider the Hadamard-classical circuit with 1-qubit (HC1Q) model~\cite{HC1Q}. 
In the HC1Q model, classical reversible gates such as $X$-gates, $CX$-gates, and TOFFOLI gates, are sandwiched between the Hadamard layers (i.e., $H^{\otimes n-1}\otimes I$). 
The reduction from our circuits to the HC1Q circuits can be understood as follows: 
In Ref~\cite{HC1Q}, 
a method to construct an HC1Q circuit from an $N$-qubit operator $U$ is introduced, where $U$ consists of Hadamard gates and classical reversible gates. The HC1Q circuit is constructed as to generate the state $U|0^N\rangle$ with postselections. 
As it is seen from our proofs, we have only used Hadamard gates and classical reversible gates except for the $Z$-gate applied to the last register. This $Z$-gate can also be implemented as $HXH$. Therefore, we can convert our circuits to HC1Q circuits using this method. Ref~\cite{HC1Q} shows that additional $h+2$ qubits are needed in this reduction, where $h$ is the number of $H$-gates used in $U$. 

Next, we think of the one-clean-qubit model (DQC1 model)~\cite{KL} and especially the case of the DQC$1_1$, in which a single output qubit is measured. 
The reduction to the DQC1$_1$ model is understood as follows: 
Although we have considered multiple-qubit-measurements, this can be easily converted into a single-qubit-measurement by changing the $X$ basis measurements into $Z$ basis measurements with $H$-gates  and then using the generalized TOFFOLI gate. 
Let us denote the acceptance probability defined through this single-qubit-measurement as $p$, which is also proportional to $gap^2$. 
We can construct DQC1$_1$ circuits whose acceptance probability (i.e. the probability of obtaining 1 when the output qubit is measured) satisfies 
\begin{eqnarray*}
\tilde{p}={4p(1-p)}/{2^N},
\end{eqnarray*}
by using the method introduced in~\cite{FKMNTT}. In this reduction, an additional qubit is needed, 
which is the clean qubit of the DQC1$_1$ model.  
Then, the same argument can be applied to the DQC1$_1$ circuits because $\tilde{p}=0$ if $p=0$ and $\tilde{p}>0$ if $0<p<1$. 

\appendix

\section{Quantum Circuit for comparing two binary integers}
\label{app:judge}
We introduce a quantum circuit which 
compares the magnitude of two binary integers. 
First, 
it is well known that the subtraction between 
two binary integers can be converted into addition 
by using 2's complement. 
When we have two $n$-bit binary integers 
$a=(a_0,...,a_{n-1})$ and $b=(b_0,...,b_{n-1})$, 
we insert a bit which represents the $\pm$ sign of them 
and define $(n+1)$-bit binary strings as 
$A\equiv(a_0,...,a_{n-1},a_n)$ and 
$B\equiv(b_0,...,b_{n-1},b_n)$. 
In this case, 
$a_n=b_n=0$ 
because both $a$ and $b$ are positive integers. 
Then the following holds:
\begin{eqnarray}
\label{eq:binarysubtract}
A-B = A + (-B) = A + B^* + 1,
\end{eqnarray}
where 
\begin{eqnarray*}
B^*\equiv(b_0\oplus 1,...,b_{n-1}\oplus 1,b_n\oplus 1)
\equiv(b_0^*,...,b_{n-1}^*,b_n^*).
\end{eqnarray*}
For example, 
when $I[a]=3$ and $I[b]=5$, 
then $a=(1,1,0)$, $b=(1,0,1)$, 
$A=(1,1,0,0)$ and $B=(1,0,1,0)$. 
Thus, $A+B^*+1=(1,1,0,0)+(0,1,0,1)+(1,0,0,0)
= (0,1,1,1,0)$, 
which correctly encodes $-2$. 

\begin{figure}[btp]
\includegraphics[width=8cm]{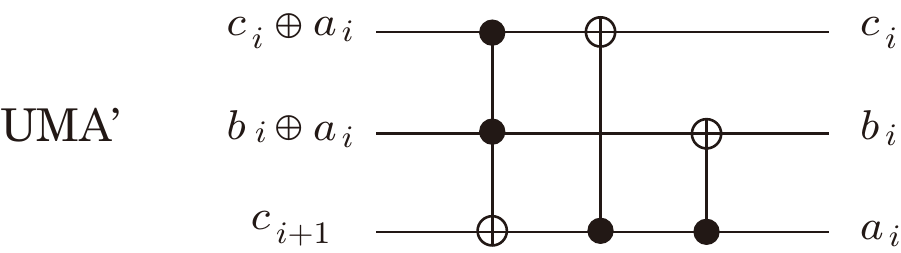}
\caption{Implementation of UMA' gate.
}
\label{fig:uma2}
\end{figure}

\begin{figure}[tbp]
\includegraphics[width=8.6cm]{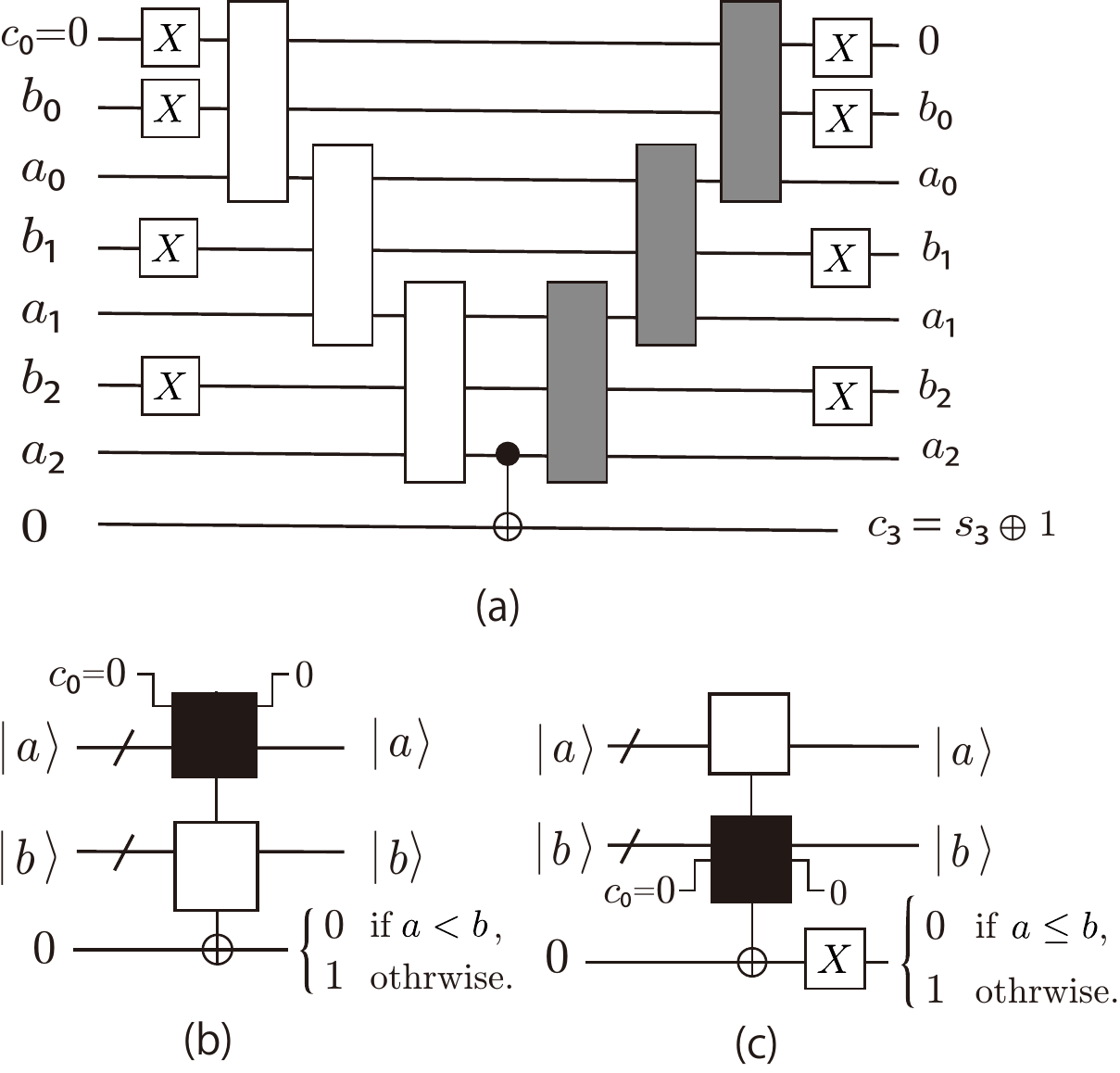}
\caption{
(a): An example of quantum circuit for $n=3$. 
The white boxes are MAJ gates and the gray boxes
are UMA' gates. (b): (a) is drawn in this way in the main text.
This is used when we want to know whether $a<b$ or $a\geq b$.  
(c): When we want to know whether $a\leq b$ or $a>b$, 
we use in this way.
}
\label{fig:compare}
\end{figure}

As it can be seen from Eq.~(\ref{eq:binarysubtract}), 
the circuit for subtraction can be implemented 
in the similar way to the addition circuit 
of Appendix~\ref{app:addition}. 
We need to change $c_0$ into $1$ 
for the added $1$ of Eq.~(\ref{eq:binarysubtract}). 
In this setting, 
$A+B^*+1$ can be written as
\begin{eqnarray*}
A+B^*+1 = (s_0,...,s_{n-1},s_n,s_{n+1}),
\end{eqnarray*}
where
$s_i=a_i\oplus b_i^*\oplus c_i$ for all $i<n+1$, 
$s_{n+1}=c_{n+1}$ 
and $c_{i+1}=MAJ(a_i,b_i^*,c_i)$ for $i>0$.
What we want to know is the $\pm$ sign of $A-B$, 
which is represented by $s_n=a_n\oplus b_n^*\oplus c_n = c_n\oplus 1$ 
and we do not need to know about the detail of $s_0,...,s_{n-1}$ and $s_{n+1}$. 
For this purpose, 
we introduce UMA' gate as Fig.~\ref{fig:uma2}, 
which just do ``UnMajority'' and do not do addition. 
For the register of $s_{n+1}$, 
we just ignore it. 
We can construct a quantum circuit which can calculate $s_n$ in this way. 
We provide an example of this circuit for $n=3$, 
which can judge whether $a<b$ or not. 
This quantum circuit is referred to as $C'$ in the main text. 
When we want to know whether $a\leq b $ or not, 
we use this circuit as (c) of Fig.~\ref{fig:compare}.
This quantum circuit is referred to as $C$ in the main text. 
The circuit $C$ uses $2n+3$ $X$-gates, $4n+1$ 
Controlled-$X$ ($CX$) -gates 
and $2n$ TOFFOLI gates.  
The circuit $C'$ uses $2n+2$ $X$-gates, $4n+1$ $CX$-gates 
and $2n$ TOFFOLI gates.

\section{Addition Circuit}
\label{app:addition}
\begin{figure}[htbp]
\begin{center}
\includegraphics[width=0.3\textwidth]{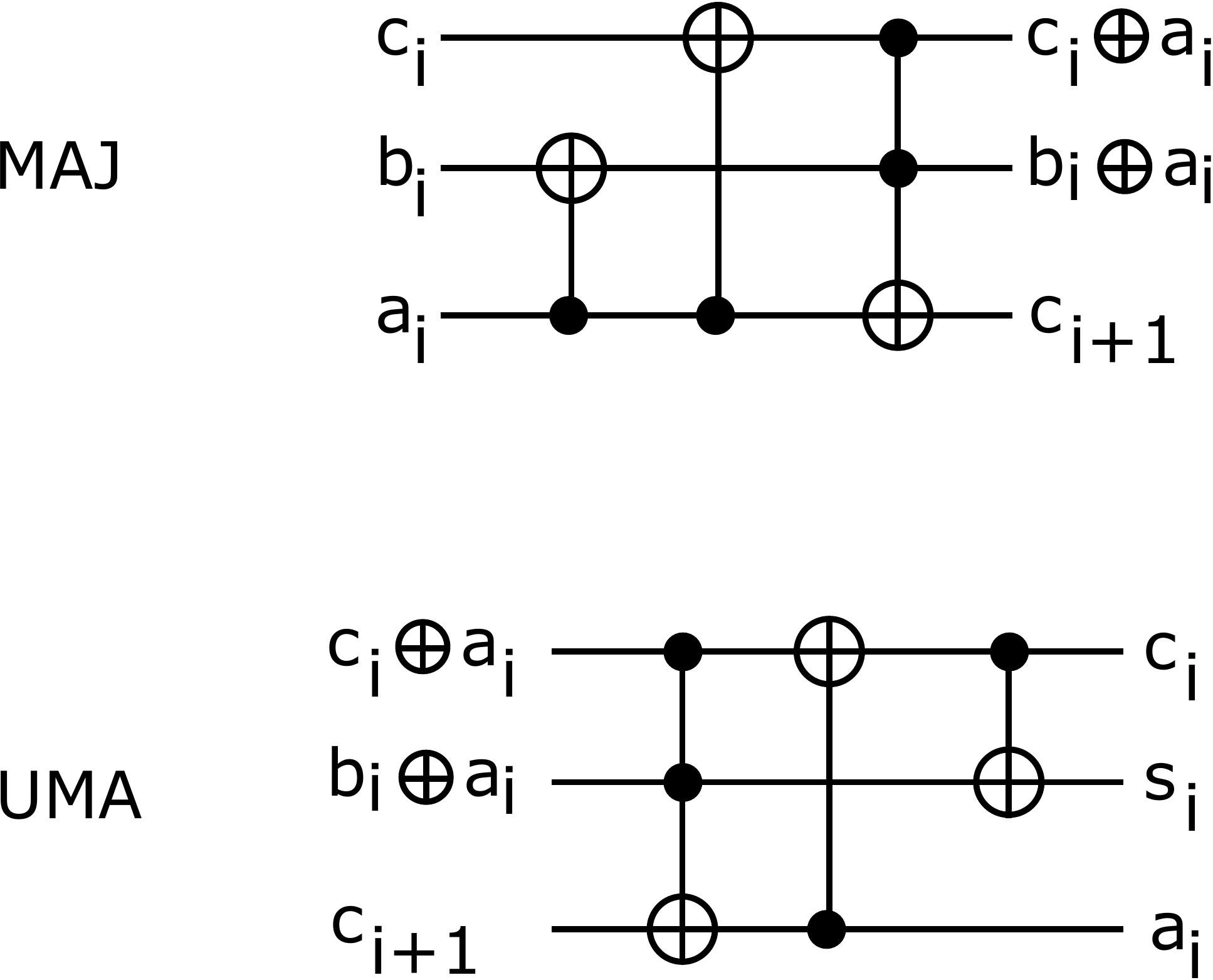}
\end{center}
\caption{The MAJ gate and the UMA gate.
}
\label{fig:MAJUMA}
\end{figure}

Here we explain the addition circuit of Ref.~\cite{addition}.
Let $a=\sum_{j=0}^{r-1}2^ja_j$ and
$b=\sum_{j=0}^{r-1}2^jb_j$ be two non-negative integers,
where $(a_0,...,a_{r-1})\in\{0,1\}^r$ and
$(b_0,...,b_{r-1})\in\{0,1\}^r$.
Let us define the MAJ gate and the UMA gate as is shown in Fig.~\ref{fig:MAJUMA}.
Here, $c_0=0$ and 
\begin{eqnarray*}
c_{i+1}=MAJ(a_i,b_i,c_i)
=a_ib_i\oplus b_ic_i\oplus c_ia_i
\end{eqnarray*}
for $i\ge0$,
and $s_i=a_i\oplus b_i\oplus c_i$ for all $i<r$ and
$s_r=c_r$.
The sum of $a$ and $b$ is $a+b=\sum_{j=0}^r 2^js_j$,
where $(s_0,...,s_r)\in\{0,1\}^{r+1}$.
This circuit uses $2n$ TOFFOLI gates and $4n+1$ $CX$-gates. 

\begin{figure}[htbp]
\begin{center}
\includegraphics[width=0.35\textwidth]{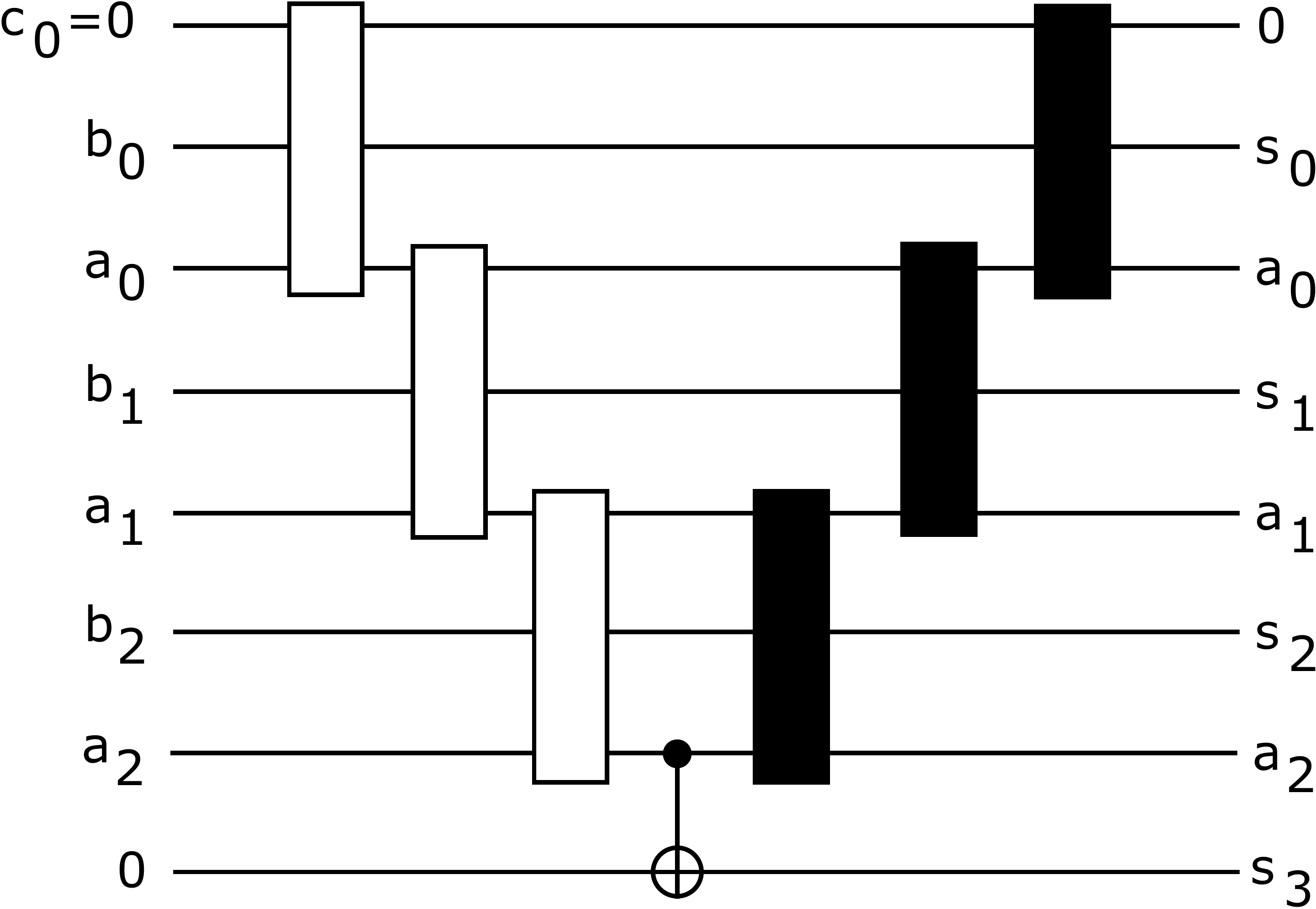}
\end{center}
\caption{An example of the addition circuit for $r=3$.
White boxes are MAJ gates, and black boxes are UMA gates.
}
\label{fig:addition}
\end{figure}

In Fig.~\ref{fig:addition}, we provide an example of the addition
circuit for $r=3$.

\begin{acknowledgments}
RH thanks Harumichi Nishimura, Francois Le Gall
and Yoshihumi Nakata for discussion.
TM thanks Ryuhei Mori for discussion.
TM is supported by MEXT Q-LEAP, JST PRESTO No.JPMJPR176A, and the
Grant-in-Aid for Young Scientists (B) No.JP17K12637 of JSPS. ST is
supported by JSPS KAKENHI Grant Numbers 16H02782, 18H04090, and
18K11164.
\end{acknowledgments}

\end{document}